\documentclass[twocolumn]{aastex631}
\usepackage{hyperref}
\usepackage{CJKutf8}

\newcommand{\lycl}{\rm{LyC\ leakers}}
\newcommand{\ionone}{\emph{Ion1}}

\newcommand{\sfsd}{$\rm \Sigma_{SFR}$}
\newcommand{\sfsdu}{$\rm log(\Sigma_{SFR}/M_{\odot}\ yr^{-1}\ kpc^{-2})$}

\begin{document}
\begin{CJK*}{UTF8}{gbsn}

\title{Lyman Continuum Leakers at $z>3$ in the GOODS-S Field: Mergers Dominated}

\author[0000-0002-2528-0761]{Shuairu Zhu (朱帅儒)}
\affiliation{Shanghai Astronomical Observatory, Chinese Academy of Sciences, 80 Nandan Road, Shanghai 200030, People’s Republic of China}
\affiliation{School of Astronomy and Space Sciences, University of Chinese Academy of Sciences, No. 19A Yuquan Road, Beijing 100049, People’s Republic of China}

\author[0000-0002-9634-2923]{Zhen-Ya Zheng*}
\affiliation{Shanghai Astronomical Observatory, Chinese Academy of Sciences, 80 Nandan Road, Shanghai 200030, People’s Republic of China}
\correspondingauthor{Zhen-Ya Zheng}
\email{*email: zhengzy@shao.ac.cn}

\author[0000-0001-6763-5869]{Fang-Ting Yuan}
\affiliation{Shanghai Astronomical Observatory, Chinese Academy of Sciences, 80 Nandan Road, Shanghai 200030, People’s Republic of China}

\author[0000-0002-0003-8557]{Chunyan Jiang}
\affiliation{Shanghai Astronomical Observatory, Chinese Academy of Sciences, 80 Nandan Road, Shanghai 200030, People’s Republic of China}

\author[0000-0003-3987-0858]{Ruqiu Lin}
\affiliation{Shanghai Astronomical Observatory, Chinese Academy of Sciences, 80 Nandan Road, Shanghai 200030, People’s Republic of China}
\affiliation{School of Astronomy and Space Sciences, University of Chinese Academy of Sciences, No. 19A Yuquan Road, Beijing 100049, People’s Republic of China}

\begin{abstract}
Understanding the ionizing photon escape from galaxies is essential for studying Cosmic Reionization. With a sample of 23 Lyman Continuum (LyC) leakers at $3<z<4.5$ in the GOODS-S field, we investigate their morphologies using high-resolution data from the Hubble Space Telescope (HST) and the James Webb Space Telescope (JWST). We find that 20 of the 23 LyC leakers show merging signatures, while the remaining 3 are starbursts. Based on our previous finding that LyC leakers are not necessarily starbursts while some are in the star formation main sequence, we further find that those in the main sequence show merger signatures. 
Our results suggest that LyC leakers are either starbursts or mergers, both of which can facilitate the LyC photon escape, in addition to generating more LyC photons.
Furthermore, we show that high-$z$ LyC leakers are statistically more extended than those selected at low redshift, which exhibits a higher merger fraction as size increases. This is likely due to the observational bias that the spatial resolution limits the detection of high-$z$ compact galaxies, while low redshift LyC leakers are more selected as compact starbursts.
\end{abstract}

\section{Introduction}
\label{sec:intro}
The Epoch of Reionization (EoR) is a crucial period in both the formation and evolution of first-generation objects and the last phase change of the whole Universe. The Gunn-Peterson troughs observed in quasars' spectra \citep[e.g.,][]{Fan2006} imply that the EoR ends at redshift $z\sim 6$. 
The CMB observations \citep[e.g.,][]{Planck2020} suggest that the EoR begins as early as $z\sim 12$. 
In the EoR, most of the hydrogen atoms in the intergalactic medium (IGM) transfer from neutral to ionized by the Lyman Continuum photons (LyC photons; $\lambda_\textrm{rest} < 912$ \AA) emitted by objects in the early Universe, such as star-forming galaxies \citep[e.g., ][]{Robertson2015} and quasars \citep[e.g., ][]{Madau2015}. 

Before the \emph{James Webb Space Telescope} (JWST) became operational, studies based on relatively bright samples of active galactic nuclei (AGN) suggested that AGNs at high redshifts played only a minor role in cosmic reionization \citep[e.g.,][]{Jiang2022, Matsuoka2023}.
Recent observations with JWST have revealed an unexpectedly abundant population of faint AGNs, which is higher than typical predictions from extrapolated AGN luminosity functions.
However, most of these faint AGNs are dust-reddened, thus resulting in low escape fractions of ionizing photons and excluding them as the primary contributors to reionization \citep[e.g.,][]{Dayal2024, Matthee2024}.
Currently, galaxies are widely believed as the most likely candidates for dominating the reionization process \citep[e.g.,][]{Finkelstein2022}.
However, contributions from galaxies to the LyC photon budget remain uncertain, because observing LyC emission from galaxies in the EoR is impossible due to the absorption by the intervening IGM \citep[e.g., ][]{Inoue2014, Robertson2022}. 

To study the escape of LyC photons from galaxies, people had to focus on galaxies with direct LyC emission detected at lower redshifts in the post-reionization era. 
To date, dozens of galaxies at $3<z<4.5$ have been identified with LyC detections \citep[e.g., ][]{Vanzella2012, Shapley2016, Vanzella2016, Yuan2021, Marques-Chaves2021, Marques-Chaves2022}.
In local Universe ($z \sim 0.3$), the Low-redshift LyC Survey \citep[LzLCS, ][]{Flury2022} and other studies \citep[e.g., ][]{Izotov2016a, Izotov2016b, Izotov2018a, Izotov2018b, Izotov2021, Wang2019, Roy2024} have identified approximately 50 galaxies as LyC leakers based on Cosmic Origins Spectrograph (COS) onboard \emph{Hubble Space Telescope (HST)}.
At $z \sim 1$, several galaxies also have been reported as LyC leakers based on the observations made by AstroSat \citep[e.g., ][]{Saha2020, Dhiwar2024, Maulick2024a, Maulick2024b}.

Based on LyC leaker samples at lower redshifts, previous research has studied the properties of LyC leaking galaxies.
These studies have also identified several observables that can trace such leakage of the LyC photons, 
for instance, high [O\,{\sc iii}]/[O\,{\sc ii}] ratios \citep[e.g., ][]{Jaskot2013, Nakajima2014}, complex Lyman-$\alpha$ line profiles \citep[e.g., ][]{Verhamme2015, Verhamme2017, Fletcher2019}, and elevated star formation surface densities \citep[e.g., ][]{Verhamme2017, Naidu2020}. 
In \citet{Zhu2024}, we found that intense bursts of star formation are not a prerequisite for the leakage of LyC photons in galaxies at $z>3$.

In addition, there is a lack of systematic studies on the galaxy morphology of LyC leakers, which may reveal the physics behind the escape of LyC photons.
In the low-$z$ studies, people are used to selecting galaxies with compact morphology for observing LyC leaking because the compactness of stellar formation and gas were thought to be the critical condition for LyC photon escape \citep[e.g., ][]{Izotov2018a, Leclercq2024}.
However, some other works propose interactions between galaxies could also facilitate the escape of LyC photons \citep[e.g., ][]{Bridge2010, Bergvall2013}.
Recently, \cite{Maulick2024a} have reported LyC detections from a merger system at $z \sim 1$.
In particular, some LyC leakers at high redshifts have also shown signs of merging activity \citep[e.g., ][]{Yuan2024, Gupta2024}.
Furthermore, the 21cm observations of a famous local LyC leaker, Haro 11, suggest that merger activity could facilitate the escape of LyC photons \citep{LeReste2023}.

These different morphological results motivate us to systematically examine the morphology of LyC leakers, taking advantage of the high spatial resolution images by the JWST and the HST. 
We have collected a sample of LyC leakers at $3<z<4.5$ in the Great Observatories' Deep Survey-South (GOODS-S), as described in the previous work \citep{Zhu2024}, where we performed a systematic study of their physical properties.
Here, we study the morphological features that could be linked to the LyC photon leakage.
Using images from the JWST and the HST, we identify mergers in our sample and measure the UV sizes of these sources.
We further check the star formation rate surface density (\sfsd) for our sample, which has been proposed as an important proxy for LyC escaping \citep[e.g.,][]{Verhamme2017, Naidu2020}.

The paper is organized as follows. We describe the sample of \lycl\ at $3<z<4.5$ and the corresponding imaging data in Section~\ref{sec:samp}.
In Section~\ref{sec:res}, we present the methods and results of morphology analysis.
We discuss properties that could connected with LyC photon escaping in Section~\ref{sec:discuss}.
Throughout this paper, we adopt a standard cosmology model with parameters $\Omega_{M} = 0.3$, $\Omega_{\Lambda} = 0.7$, and $H_{0} = \rm{70\ km\ s^{-1} Mpc^{-1}}$. 
All magnitudes here are given in the AB system.

\section{Sample and Data}
\label{sec:samp}
We have collected a sample of LyC leakers at $3<z<4.5$ (from 3.084 to 4.426, with an average value of 3.505) and have examined their UV-to-IR spectral energy distributions (SED) in our previous work \citep{Zhu2024}, where the details of the sample can be found.
The sample is a collection of LyC leakers in the GOODS-S field from the literature \citep{Ji2020, Debarros2016, Yuan2021, Saxena2022, Rivera-thorsen2022, Gupta2024, Kerutt2024}.
The sample contains 23 LyC leakers after excluding potential contaminants and redundant objects. 
All their LyC emission is detected at more than a 2$\sigma$ level in at least one observation.
Eight are considered high-confidence LyC leakers, whose LyC emissions are detected at a higher than $3 \sigma$ level.
We have analyzed their physical properties in \citet{Zhu2024} and compared their star formation with the star formation main sequence (SFMS).
We find that not all the LyC leakers are undergoing an intense starburst.

To keep consistency with \citet{Zhu2024}, we also include high-confidence LyC leakers in other fields for comparison.
\citet{Zhu2024} include three LyC leakers in other fields, which are \emph{Ion3} at $z \sim 4$, J0121+0025 at $z \sim 3.244$, and J1316+2614 at $z \sim 3.6130$ \citep{Vanzella2018, Marques-Chaves2021, Marques-Chaves2022, Marques-Chaves2024}.
However, \emph{Ion3} and J0121+0025 lack HST coverage and are only marginally resolved in ground-based images, limiting detailed morphological analysis. 
Only J1316+2614 has been recently observed with HST \citep{Marques-Chaves2024}, so we focus on this LyC leaker in this study.

We analyze the morphology of LyC leakers with images from JWST and HST. We utilize the final combined JWST Near-Infrared Camera (NIRCam) images made public by the JADES Data Release 2, which include a series of broad and medium bands \citep{Eisenstein2023b}.
These final images comprise exposures from the JWST Advanced Deep Extragalactic Survey \citep{Eisenstein2023}, the First Reionization Epoch Spectroscopically Complete Observations \citep[FRESCO, ][]{Oesch2023}, and the JWST Extragalactic Medium-band Survey \citep[JEMS, ][]{Williams2023}.
Additionally, we use the HST images released by the Hubble Legacy Field \citep[HLF, ][]{Illingworth2016, Whitaker2019}, which has combined images taken by HST over 18 years in this field.
The HST images are available for all 23 sources, while the JWST images are available for 16.

\section{Morphological analysis}
\label{sec:res}
In this section, we perform a morphological analysis of these LyC leakers in the GOODS-S.
The structures of the sources are well-resolved by HST or JWST. 
Our merger identification and size measurement methods and results are presented in Section~\ref{subsec: morph} and Section~\ref{subsec:size}, respectively.

\subsection{Merger Identification}
\label{subsec: morph}
\begin{figure*}[tb]
    \centering
    \includegraphics[width=0.85\textwidth]{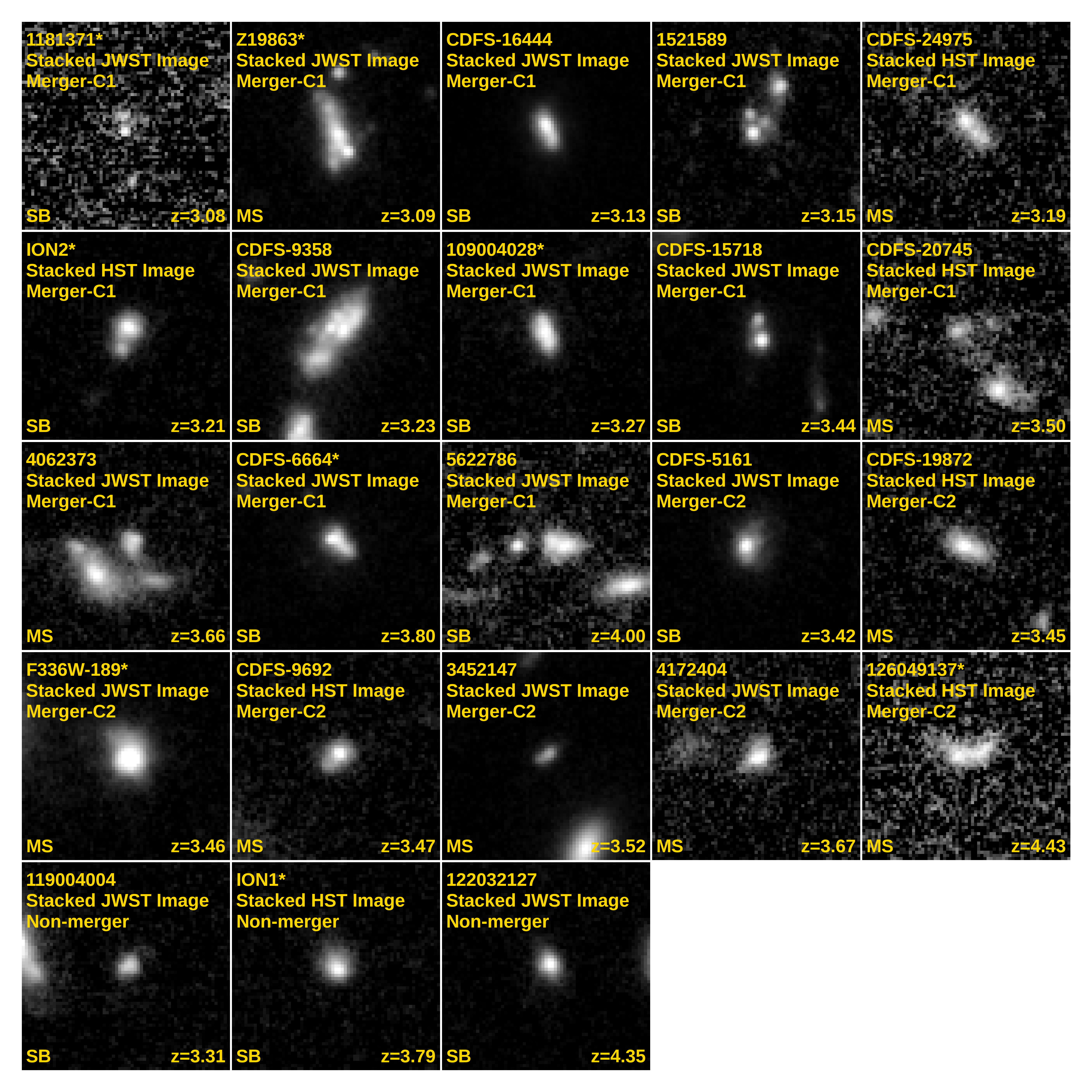}
    \caption{
    High-resolution $2 \arcsec \times 2 \arcsec$ cutouts for the \lycl. In the top-left corner of each cutout, we show the source ID, the type of image used, and the criteria for identifying mergers.
    High-confidence LyC leakers are marked with a star next to their IDs. The redshift is indicated in the bottom-left corner and the star formation activity is shown in the bottom-right corner (MS for the star-forming main sequence and SB for starbursts).}
    \label{fig:cutouts}
\end{figure*}

Merger identification relies on the highest-resolution images available for each source. 
We visually inspect the stacked JWST and HST images for each source.
The stacked JWST image comprises the F182M, F200W, and F210M images, and the stacked HST image comprises the F775W, F814W, and F850LP images. 
Following \citet{Whitaker2019}, the stacked image is a noise-equalized image combining science images by multiplying the square root of the inverse variance maps.

We follow \citet{Jiang2013} on the classification criteria for mergers.
Two types of galaxy mergers are identified in this work: (1) galaxies with two or more cores and/or close companions and (2) galaxies that exhibit extended, elongated structures or long tails.
Out of the 23 LyC leakers in the sample, we identify 20 mergers, with 13 based on the first criterion and 7 based on the second criterion (See Fig.~\ref{fig:cutouts}).
According to the above criteria, the merger fraction is $\sim$ 87\% (20/23).
Among these merging LyC leakers, 10 are starbursts, and 10 are on the star-forming main sequence \citep{Zhu2024}.
In fact, all the LyC leakers on the star-forming main sequence belong to merging LyC leakers.
The remaining three galaxies which do not pass the above criteria also show some irregularities.
However, these irregularities are not significant, so we do not count them as mergers in this work.
Two exhibit offsets between the LyC and the non-ionizing emission, including one high-confidence LyC leaker \ionone\ \citep{Ji2020}.
This offset may indicate that the ISM of the object is disturbed in a state of merging \citep{Yuan2024}. Limited sensitivity or projection effects may obscure merging features in the images.

Besides the visual inspection, we also test the quantitative methods in identifying mergers, such as the nonparametric morphological method (e.g., CAS and Gini; \citealt{Conselice2003, Lotz2004}). 
We then apply criteria from previous studies \citep[e.g., ][]{Conselice2003, Lotz2008b} to identify mergers.
Only eight of our galaxies can be identified as mergers, including 7 mergers identified by the above visual method, and \ionone\, which does not show significant merger signatures in the image.
The other 13 mergers identified by visual inspection are not identified as mergers using the nonparametric method.
This may be due to a lack of resolution for these methods to work well.
As shown in previous works, for high-$z$ galaxies, these parameters can be biased and unreliable \citep[e.g., ][]{Jiang2013}. 
Besides, they are only sensitive to specific merger stages, potentially misclassifying non-merging systems as mergers in the rest-frame UV images \citep[e.g., ][]{Kartaltepe2010, Snyder2019, Rose2023}.

\subsection{Size Measurement}
\label{subsec:size}
We use the half-light radii ($r_{50}$) to describe the size of high-$z$ LyC leakers.
We measure $r_{50}$ of these LyC leakers based on the HLF ACS/F814W image, which probes the rest-frame UV wavelength from 1530 \AA\ to 2140 \AA\ at $z \sim 3.5$. We first create $9^{\arcsec} \times 9^{\arcsec}$ cutouts and subtract the background for each source. 
The background means and noise are estimated using sigma-clipped statistics after masking sources in the images. 
A segmentation image is created to mask nearby sources during the measurements.
The half-light radius of each source is measured using \texttt{SExtractor} \citep{1996A&AS..117..393B}.
To estimate the uncertainty of the half-light radius, we randomly move the sources of interest 320-360 times in nearby regions and measure the half-light radius each time using the same method.
To correct the effect of the point spread function (PSF) on the half-light radius, we use the formula,
\begin{equation}
\label{eq:psf}
    r_{50} = \sqrt{r_{50, app}^{2} - r_{50, \rm PSF}^{2}},
\end{equation}
where $r_{50, app}$ is the apparent value we measured, $r_{50, \rm PSF}$ is $r_{50}$ represents the PSF's half-light radius in the ACS/WFC F814W band, and $r_{50}$ is the half-light radius after correcting the PSF effect.
The uncertainty is then estimated based on the standard deviation of these measurements.

\begin{figure*}[t]
    \centering
    \includegraphics[width=0.85\textwidth]{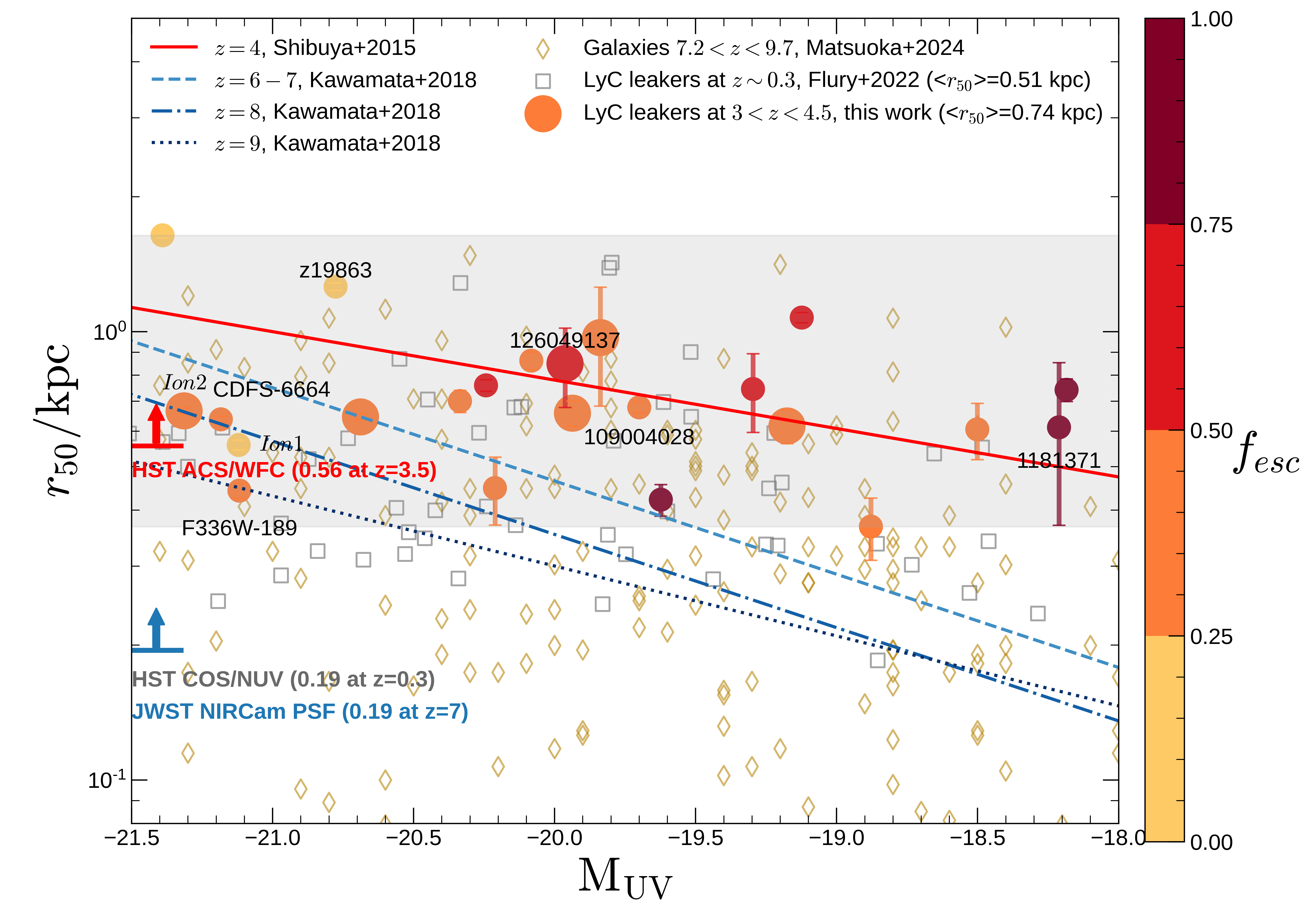}
    \caption{Half-light radius ($r_{50}$) versus UV absolute magnitude $\mathrm{M_{UV}}$ for high-$z$ ($3<z<4.5$) LyC leakers (dots) in the GOODS-S.
    The sizes ($r_{50}$) are derived from the ACS/F814W image.
    We highlight LyC leakers where the LyC signal is not offset from the non-ionizing emission using larger dots.
    We also mark high-confidence LyC leakers by their names.
    The color bar encodes the $f_{\rm esc}$ values of the LyC leakers.
    The shaded region shows the size range of the sample that we use to select the comparison samples at other redshifts.
    For comparison, low-$z$ LyC leakers \citep{Izotov2016a, Izotov2016b, Izotov2018a, Izotov2018b, Wang2019, Flury2022} and galaxies at $7.2<z<9.7$ selected as NIRCam/F090W dropouts \citep{Morishita2024} are shown as squares and diamonds, respectively.
    The size-$\rm M_{UV}$ relations for LBGs at $z \sim 4$, $z \sim 6.5$, $z \sim 8$, and $z \sim 9$ are also displayed by blue lines \citep{Shibuya2015, Kawamata2018}.
    Blue arrows denote $r_{50}$ of the PSF for ACS/WFC F814W, COS/NUV, and JWST NIRCam/F150W, and the values are 0.56 kpc at $z=3.5$, 0.19 kpc at $z=0.3$, and 0.19 kpc at $z=7$, respectively.
    }
    \label{fig:re_muv}
\end{figure*}

In Fig.~\ref{fig:re_muv}, we present $r_{50}$ as a function of UV absolute magnitude for \lycl\ in our sample, with circles color encoded by the escape fraction.
The sizes of LyC leakers in our sample range from $r_{50}\sim$ 0.37 kpc to 1.64 kpc with a mean $r_{50}$ value of 0.74 kpc.

No significant size difference is found between starburst LyC leakers (a mean value of 0.72 kpc) and main sequence LyC leakers (a mean value of 0.76 kpc).
There is no clear dependence of the sizes on the UV magnitude for the high-$z$ leakers, although previous works find the size and $\rm M_{UV}$ are related for Lyman Break Galaxies (LBGs) at $z \sim 4-9$ \citep[e.g., ][]{Shibuya2015, Kawamata2018}.
This may be due to the limited spatial resolution, together with the insufficient depth of the current UV images (See Section~\ref{subsec:discuss-selectionbias} for more details).
We do not find any dependence of sizes on $f_{\rm esc}$ for our sample.
Furthermore, we divide the high-$z$ LyC leakers into two groups according to whether they show offsets between LyC and non-ionizing UV emission.
We find no systematic size difference between galaxies with and without LyC offsets.

\section{Discussion}
\label{sec:discuss}

\subsection{Mergers in high-$z$ LyC leakers}
\label{subsec:discuss-merger}
The fraction of mergers in our sample of high-$z$ LyC leakers in the GOODS-S is nearly 87\% (20/23).
This is significantly higher than the merger fraction at similar or higher redshifts. 
\citet{Jiang2013} derived a fraction of 40\%-50\% based on bright galaxies in their sample of Ly$\alpha$ emitters and LBGs at $z \sim 6$. They concluded the fraction would be lower if faint galaxies ($\rm M_{UV} \gtrsim -20.5$) were included because the merger systems usually have stronger UV emission and star formation activity.

The unusually high fraction of mergers in our sample suggests that merger activity may facilitate the escape of LyC photons. The merging and interacting processes of galaxies can disturb the interstellar medium (ISM), creating channels with lower optical depth that allow LyC photons to escape \citep{Rauch2011, Gupta2024, Yuan2024, Maulick2024a}.
A recent 21 cm observation by \citet{LeReste2023} of a nearby LyC leaker (Halo 11) also supports this scenario. They found that merger-driven interactions can displace the bulk of neutral gas from LyC emission sources, facilitating the escape of these photons.
Interestingly, \citet{Witten2024} used JWST to study a sample of Ly$\alpha$ emitters at $z>7$ and found that all emitters have close companions.
Furthermore, they showed that merger activity could facilitate the escape of Ly$\alpha$ photons by clearing up neutral gas in several channels based on simulations. LyC photons can also escape through (part of) these channels.

Among the three LyC leakers in our sample identified as non-mergers, source 119004004 shows no significant offset between LyC and non-ionizing UV emission. This LyC leaker is undergoing a vigorous starburst with a specific star formation rate (sSFR) of $\log (\mathrm{sSFR}$/yr) $\simeq-6.95$.
The other two LyC leakers, \ionone\ and 122032127, which display significant offsets between LyC and non-ionizing UV emission and irregular shapes, also host intense star formation activity.
Generally, LyC leakers in our sample classified as non-mergers have more intense star formation than those of mergers, with mean sSFR $\log (\mathrm{sSFR}$/yr) $\simeq$ -6.95 and -7.65, respectively.
For the LyC leaker J1316+2614 which is not in the GOODS-S field, it exhibits neither merger signatures nor significant offsets but has intense star formation, too.
In such intense star formation activities, gas turbulence and outflows driven by radiative and mechanical feedback can also create channels for the escape of LyC photons \citep{Amorn2024}. It appears that if the production and escape of LyC photons are not facilitated by mergers, intense star formation has to take place to ensure observable LyC photons.

We also examine the merger fraction in the sample of LyC leakers at $z \sim 0.3$ to assess whether merger activity facilitates LyC photon escape in low-$z$ systems.
By examining the HST COS/NUV images of these $z \sim 0.3$ LyC leakers, we find that half of LyC leakers also exhibit merger signatures (24/50), and almost all the mergers have similar sizes to ours (20/30).

It appears that merging becomes more dominant as LyC leakers become more extended. For small galaxies with more efficient stellar feedback, if star-forming activities are strong enough, considerable LyC photons can be produced, and a fraction of them may escape into the IGM. These small galaxies would show high \sfsd, which is a usually adopted criterion for selecting low-$z$ LyC leaker candidates \citep[e.g., ][]{Flury2022}. In our sample, LyC leakers of non-mergers also have more intense star formation than mergers. On the other hand, LyC leakers on the star formation main sequence at $3<z<4.5$ are all mergers \citep{Zhu2024}. It suggests that LyC photon escape happens in intense starbursts or mergers and interactions.

\subsection{Sizes of LyC leakers}
\label{subsec:discuss-size}
In this section, we discuss the sizes of LyC leakers at high-$z$ ($3<z<4.5$), in comparison with the sizes of those at low-$z$ ($z\sim$ 0.3), and the sizes of star-forming galaxies at $z>5$ measured by JWST.

In Fig.~\ref{fig:re_muv}, we show the size $r_{50}$ as a function of absolute UV magnitude $\rm M_{UV}$, with circles for LyC leakers at $3<z<4.5$.
For low-z LyC leakers, we include 50 galaxies from previous studies \citep[squares, ][]{Izotov2016a, Izotov2016b, Izotov2018a, Izotov2018b, Wang2019, Flury2022}.
We use $\rm M_{UV}$ and $r_{50}$ values from \citet{Flury2022}, where $\rm M_{UV}$ is derived from the HST COS spectrum and $r_{50}$ is measured from HST COS/NUV acquisition images.
The $\rm M_{UV}$ is based on the best-fit spectrum, and the $r_{50}$ is determined from light growth curves measured using a series of apertures with increasing radii until total flux is reached.
We note that  \citet{Flury2022} did not correct for PSF effects in $r_{50}$, so we apply this correction using Formula~\ref{eq:psf}.
We also consider star-forming galaxies at $z>5$ for which the sizes are measured using \texttt{Galfit} \citep{Peng2010} based on JWST images \citep{Morishita2024}, including those at $5.0 < z < 7.2$ (F070W dropouts), $7.2 < z < 9.7$ (F090W dropouts), and $9.7 < z < 13.0$ (F115W dropouts).
For clarity, we only display the sizes of those at $7.2<z<9.7$ in Fig.~\ref{fig:re_muv}.

We find that LyC leakers at $z \sim 0.3$ and star-forming galaxies at $z>5$ are statistically more compact than our LyC leakers, with mean sizes of 0.51 kpc for LyC leakers at $z \sim 0.3$ and 0.39 kpc, 0.39 kpc, and 0.33 kpc for F070W, F090W, and F115W dropouts, respectively.
In contrast, the LyC leakers in our sample are more extended with a mean size of $r_{50} = 0.74$ kpc.
The size is a critical factor in determining the physical properties of galaxies \citep[e.g., ][]{Malhotra2012}.
The mechanisms driving LyC photon escape may differ across systems of varying sizes. 
As found in Section~\ref{subsec:discuss-merger}, more extended LyC leakers have a higher probability of being in a merging event, while more compact LyC leakers have a lower merger fraction but exhibit more intense star formation. LyC leakers at $3<z<4.5$ in GOODS-S are generally very extended and predominately mergers. LyC photon escape is likely made easy during the merging process.

\subsection{Observational bias on the size measurement of LyC leakers at high-$z$}
\label{subsec:discuss-selectionbias}
The lack of compact high-$z$ LyC leakers in our sample is more likely due to the observational bias of the limited spatial resolution and the insufficient depth of UV images in probing the LyC signals.
As shown in Fig. \ref{fig:re_muv}, the $r_{50,\mathrm{PSF}}$ values for the PSF of HST COS/NUV, HST ACS/WFC F814W, and JWST NIRCam F150W are 0\arcsec.04, 0\arcsec.08, and 0\arcsec.04, corresponding to physical scales of 0.19 kpc at $z \sim 0.3$, 0.56 kpc at $z \sim 3.5$, and 0.19 kpc at $z \sim 7$, respectively. The larger PSF size at $z \sim 3.5$ prevents the detection of more compact galaxies compared to those at other redshifts. 

The insufficient depth of UV images in probing the LyC signals at $z \sim 3.5$ would also bias the selection in favor of bright LyC galaxies, which are more likely to be extended according to the size-\(\rm M_{UV}\) relation at \(z \sim 4\) \citep[e.g.,][]{Shibuya2015}. With the current UV depth, faint LyC galaxies are detectable only when the escape fraction of LyC photons is sufficiently high. According to the estimation of the escape fraction of LyC photons \citep[see Equation 2 of ][]{Yuan2021}, the relative escape fraction of LyC photons from a galaxy is:
\begin{equation}
    f_{esc} = \frac{(L_{1500}/L_{\mathrm{LyC}})_{\mathrm{int}}}{(L_{1500}/L_{\mathrm{LyC}})_{\mathrm{obs}}} \times \frac{1}{T_{\mathrm{IGM}}}
\end{equation}
where $(L_{1500}/L_{\mathrm{LyC}})_{\mathrm{int}}$ and $(L_{1500}/L_{\mathrm{LyC}})_{\mathrm{obs}}$ are the intrinsic and observed ratios of the flux density at rest-frame wavelengths 1500 \AA\  and LyC regime, respectively, and $T_{\mathrm{IGM}}$ is the IGM transmission coefficient along a given line of sight.
Assuming the most transparent line-of-sight at $z \sim 3.5$ \citep[$T_{\mathrm{IGM}}=0.557$, ][]{Steidel2018} and a high intrinsic UV-to-LyC luminosity ratio \citep[$(L_{1500}/L_{\mathrm{LyC}})_{\mathrm{int}}=1.37$, ][]{Kerutt2024}, the depth of the UV image (HST WFC/UVIS F336W, 28 mag) from the Hubble Deep UV Legacy Survey \citep{Oesch2018} allows us to detect LyC emission from galaxies with $\rm M_{UV} = -18$ only when the escape fraction exceeds $\sim 80\%$.

\begin{figure*}[htb]
    \centering
    \includegraphics[width=0.85\textwidth]{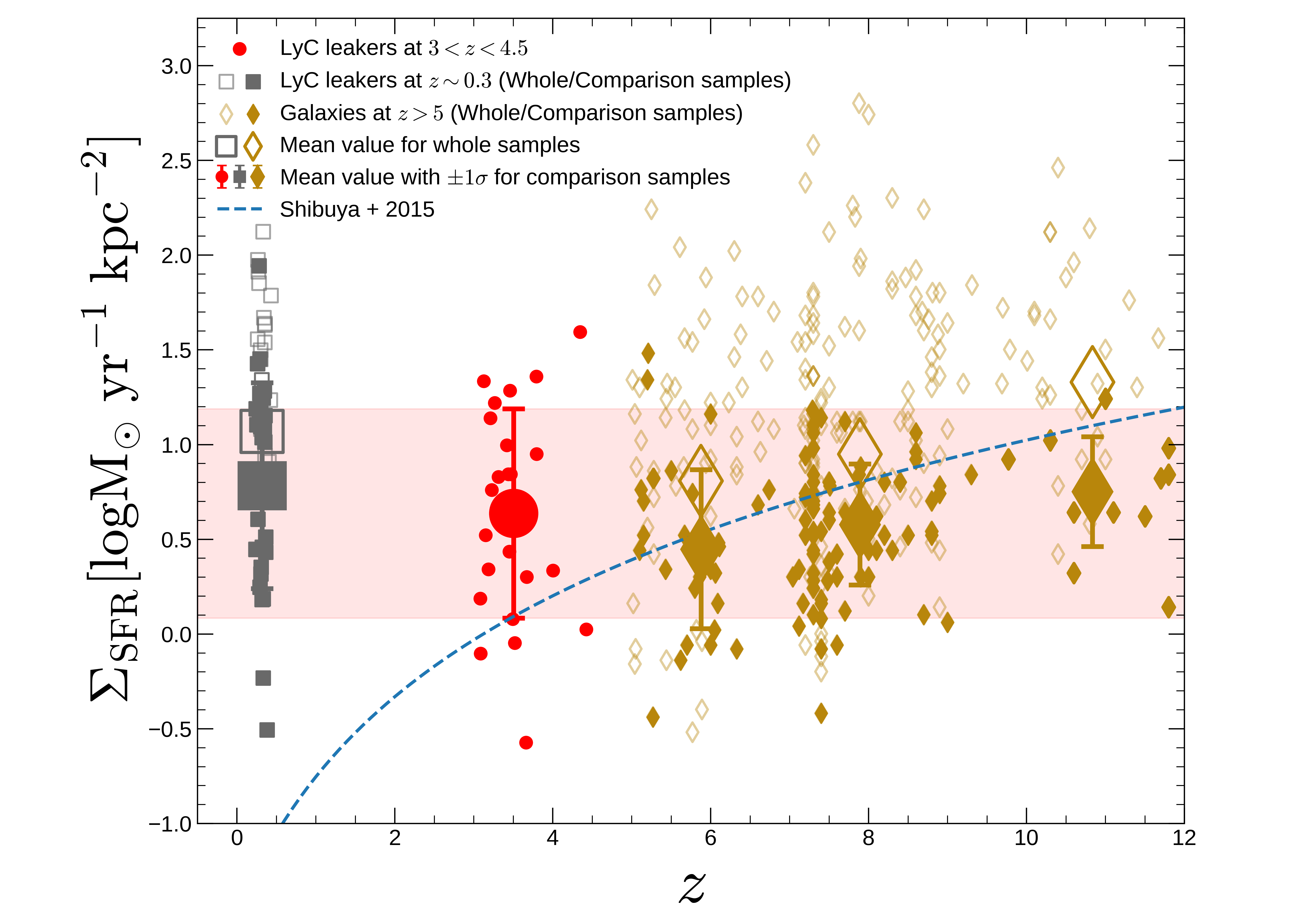}
    \caption{\sfsd\ as a function of redshift.
    Red dots represent our sample of \lycl\ at $3<z<4.5$ in the GOODS-S field, while gray squares indicate \lycl\ at $z\sim 0.3$ \citep{Izotov2016a, Izotov2016b, Izotov2018a, Izotov2018b, Wang2019, Flury2022}.
    Star-forming galaxies at $z>5$ including those at $5.0 < z < 7.2$ (F070W dropouts), $7.2 < z < 9.7$ (F090W dropouts), and $9.7 < z < 13.0$ (F115W dropouts) from \citet{Morishita2024}, are shown as gold diamonds.
    We highlight sources of comparable size and $\rm M_{UV}$ (0.5 kpc $<r_{50}<$ 1.2 kpc and $-21.5<\rm M_{UV}<-18$) to LyC leakers in our sample using filled symbols.
    The filled and open large symbols represent the mean values of the comparison samples and the whole samples, respectively, and the error bars represent 1$\sigma$ distributions for the comparison samples.
    The blue dashed line shows the \sfsd\ evolution calculated from the size evolution given by \citet{Shibuya2015}, assuming a SFR of 10 $\rm M_{\odot}\ yr^{-1}$.
    We use a red-shaded region to highlight the range of 1$\sigma$ around the mean of our sample.}
    \label{fig:sfsd}
\end{figure*}

\subsection{Star formation rate surface density}
\label{subsec:discuss-sfsd}
In \citet{Zhu2024}, we have analyzed the star formation properties of these 23 LyC leakers in the GOODS-S. We find that not all of them are starburst in which systems strong radiative and mechanical feedback could facilitate LyC photons escaping \citep[e.g., ][]{Amorn2024}.
In this section, we further examine \sfsd\ of these galaxies and compare them with LyC leakers at $z \sim 0.3$ \citep{Izotov2016a, Izotov2016b, Izotov2018a, Izotov2018b, Wang2019, Flury2022} and star-forming galaxies at $z>5$ \citep[][]{Morishita2024}.
We calculate \sfsd\ by 
\begin{equation}
\label{eq:sfsd}
    \mathrm{\Sigma_{SFR}} = \frac{\mathrm{SFR}}{2 \pi r_{50}^{2}},
\end{equation}
where SFR represents the star formation rate of galaxies (SFR). For our sample, this property of each galaxy has been derived from SED fitting in \citet{Zhu2024}.
We calculate \sfsd\ of LyC leakers at $z\sim 0.3$ using the PSF-corrected $r_{50}$ and SFR from \citet{Flury2022} which are derived from H$\beta$ luminosity.
We take \sfsd\ of galaxies at $z \gtrsim 5$ from \citet{Morishita2024} where the SFR is estimated from UV luminosity and the size of each galaxy is calculated with the JWST images.

In Fig.~\ref{fig:sfsd}, we present \sfsd\ of LyC leakers at $3<z<4.5$, in comparison with LyC leakers at $z \sim 0.3$ as well as star-forming galaxies at $z>5$.
We also present the \sfsd\ evolution of galaxies assuming a typical SFR of 10 $\rm M_{\odot}\ yr^{-1}$ \citep[][]{Shibuya2015}.
At both $z\sim 0.3$ and $z>5$, many galaxies exhibit higher \sfsd\ than our sample.
The \sfsd\ values of LyC leakers at $3<z<4.5$ span a range of \sfsdu $\simeq -0.57-1.59$, with a mean value of \sfsdu $\simeq$ 0.64.
The mean \sfsd\ of LyC leakers at $z \sim 0.3$ and the mean \sfsd\ of star-forming galaxies at $z>5$ are 0.17–0.69 dex higher than our sample.

As we mentioned in Section \ref{subsec:discuss-size}, different sizes of galaxies may be connected to different escape mechanisms of LyC photons. To investigate this, we compare galaxies and LyC leakers across different redshifts with similar sizes. Specifically, we construct comparison samples for galaxies at \( z > 5 \) and LyC leakers at $z \sim 0.3$, focusing on sources with sizes comparable to those in our sample.
The selection criteria for comparison samples, covering the size and UV magnitude range of our high-$z$ LyC leakers in the GOODS-S, are 0.37 to 1.64 kpc for $r_{50}$ and -21.5 to -18 for $\rm M_{UV}$. This results in 30 LyC leakers at $z \sim 0.3$, along with 33 F070W dropouts, 71 F090W dropouts, and 12 F115W dropouts in the selected range.

We compare \sfsd\ of our sample with the comparison samples.
We find that the mean \sfsd\ of these comparison samples are consistent with the mean \sfsd\ of LyC leakers at $3<z<4.5$ in GOODS-S within 1$\sigma$, as highlighted by the red-shaded region in Fig.~\ref{fig:sfsd}. 
The star formation rate surface density of LyC leakers with comparable sizes shows little evolution with redshift from $z \sim 0.3$ to $z \sim3.5$, and closely resembles that of star-forming galaxies at $z>5$ with similar sizes, indicating possible similar star formation properties and escape mechanism of LyC photons.

\section{Summary}
\label{sec:sum}
In this paper, we investigate the merger fraction, sizes, and star formation surface densities of LyC leakers at $3<z<4.5$ in the GOODS-S field, using high-resolution data from HST and JWST. We compare our sample to LyC leakers at $z \sim 0.3$ and star-forming galaxies at $z>5$.

We find that mergers dominate our sample of LyC leakers at $3<z<4.5$, with 20 out of 23 LyC leakers exhibiting multiple components or companions.
This results in an overall merger fraction of $\gtrsim$86\% in our sample of LyC leakers at $3<z<4.5$, notably higher than the 50\%-60\% observed in other galaxy samples at high redshifts.
In addition, all main-sequence LyC leakers in our sample are mergers, while all non-mergers are starbursts with more intense star formation than the mergers. A similar trend is observed in the low-$z$ LyC leaker sample, suggesting that LyC photon escape is driven by either intense starbursts or mergering interactions.

We measure the sizes of these LyC leakers using a model-independent method with \texttt{SExtractor} and correct for the PSF effect.
The sizes of LyC leakers at $3 < z < 4.5$ in GOODS-S range from 0.37 kpc to 1.64 kpc, with a mean size of 0.74 kpc.
We find that LyC leakers at $3<z<4.5$ in the GOODS-S are generally more extended than LyC leakers at $z \sim 0.3$ and star-forming galaxies at $z>5$, and have a much lower mean \sfsd.
This difference is caused by the observational bias due to the limited spatial resolution and the insufficient depth of UV images in probing the LyC signals. In a comparable size range, we find that our LyC leakers at $3<z<4.5$ show consistent mean values of \sfsd\ with LyC leakers at $z\sim 0.3$ and star-forming galaxies at $z>5$. 

We note that high-resolution UV images with deep depth are crucial in studying high-$z$ LyC leakers. The upcoming Multi-Channel Imager (MCI) onboard the China Space Station Telescope \citep[CSST,][]{zhan2018} will provide images across multiple bands with a resolution comparable to that of the HST, especially deeper depth and larger area in UV bands. This advancement will not only help us identify more LyC leakers at high redshifts but also enable more detailed studies of their morphology and physics.

\begin{acknowledgments}
This work is supported by National Key R\&D Program of China No.2022YFF0503402. ZYZ acknowledges support from the National Science Foundation of China (12022303) and the China-Chile Joint Research Fund (CCJRF No. 1906). FTY acknowledges support from the Natural Science Foundation of Shanghai (Project Number: 21ZR1474300). We also acknowledge the science research grants from the China Manned Space Project, especially NO. CMS-CSST-2021-A04, CMS-CSST-2021-A07. 

This work is based on observations made with the NASA/ESA/CSA James Webb Space Telescope. The data were obtained from the Mikulski Archive for Space Telescopes at the Space Telescope Science Institute, which is operated by the Association of Universities for Research in Astronomy, Inc., under NASA contract NAS 5-03127 for JWST. 
This work is based on observations taken by the 3D-HST Treasury Program (GO 12177 and 12328) with the NASA/ESA HST, which is operated by the Association of Universities for Research in Astronomy, Inc., under NASA contract NAS5-26555.

All the data used in this paper can be found in MAST: \dataset[10.17909/T9JW9Z \citep{https://doi.org/10.17909/t9jw9z}]{http://dx.doi.org/10.17909/T9JW9Z}, \dataset[10.17909/8tdj-8n28 \citep{https://doi.org/10.17909/8tdj-8n28}]{http://dx.doi.org/10.17909/8tdj-8n28} and 
\dataset[10.17909/fsc4-dt61 \citep{https://doi.org/10.17909/fsc4-dt61}]{https://doi.org/10.17909/fsc4-dt61}.
\end{acknowledgments}

\facilities{HST (ACS), JWST (NIRCam)}

\software{Astropy \citep{2013A&A...558A..33A,2018AJ....156..123A,2022ApJ...935..167A}, Numpy \citep{harris2020array}, Matplotlib \citep{Hunter:2007}, \texttt{SExtractor} \citep{1996A&AS..117..393B},
\texttt{statmorph} \citep{2019MNRAS.483.4140R}}

\bibliography{reference}{}

\begin{thebibliography}{}
\expandafter\ifx\csname natexlab\endcsname\relax\def\natexlab#1{#1}\fi
\providecommand{\url}[1]{\href{#1}{#1}}
\providecommand{\dodoi}[1]{doi:~\href{http://doi.org/#1}{\nolinkurl{#1}}}
\providecommand{\doeprint}[1]{\href{http://ascl.net/#1}{\nolinkurl{http://ascl.net/#1}}}
\providecommand{\doarXiv}[1]{\href{https://arxiv.org/abs/#1}{\nolinkurl{https://arxiv.org/abs/#1}}}

\bibitem[{{Amor{\'\i}n} {et~al.}(2024){Amor{\'\i}n}, {Rodr{\'\i}guez-Henr{\'\i}quez}, {Fern{\'a}ndez}, {V{\'\i}lchez}, {Marques-Chaves}, {Schaerer}, {Izotov}, {Firpo}, {Guseva}, {Jaskot}, {Komarova}, {Mu{\~n}oz-Vergara}, {Oey}, {Bait}, {Carr}, {Chisholm}, {Ferguson}, {Flury}, {Giavalisco}, {Hayes}, {Henry}, {Ji}, {King}, {Leclercq}, {{\"O}stlin}, {Pentericci}, {Saldana-Lopez}, {Thuan}, {Trebitsch}, {Wang}, {Worseck}, \& {Xu}}]{Amorn2024}
{Amor{\'\i}n}, R.~O., {Rodr{\'\i}guez-Henr{\'\i}quez}, M., {Fern{\'a}ndez}, V., {et~al.} 2024, \aap, 682, L25, \dodoi{10.1051/0004-6361/202449175}

\bibitem[{{Astropy Collaboration} {et~al.}(2013){Astropy Collaboration}, {Robitaille}, {Tollerud}, {Greenfield}, {Droettboom}, {Bray}, {Aldcroft}, {Davis}, {Ginsburg}, {Price-Whelan}, {Kerzendorf}, {Conley}, {Crighton}, {Barbary}, {Muna}, {Ferguson}, {Grollier}, {Parikh}, {Nair}, {Unther}, {Deil}, {Woillez}, {Conseil}, {Kramer}, {Turner}, {Singer}, {Fox}, {Weaver}, {Zabalza}, {Edwards}, {Azalee Bostroem}, {Burke}, {Casey}, {Crawford}, {Dencheva}, {Ely}, {Jenness}, {Labrie}, {Lim}, {Pierfederici}, {Pontzen}, {Ptak}, {Refsdal}, {Servillat}, \& {Streicher}}]{2013A&A...558A..33A}
{Astropy Collaboration}, {Robitaille}, T.~P., {Tollerud}, E.~J., {et~al.} 2013, \aap, 558, A33, \dodoi{10.1051/0004-6361/201322068}

\bibitem[{{Astropy Collaboration} {et~al.}(2018){Astropy Collaboration}, {Price-Whelan}, {Sip{\H{o}}cz}, {G{\"u}nther}, {Lim}, {Crawford}, {Conseil}, {Shupe}, {Craig}, {Dencheva}, {Ginsburg}, {VanderPlas}, {Bradley}, {P{\'e}rez-Su{\'a}rez}, {de Val-Borro}, {Aldcroft}, {Cruz}, {Robitaille}, {Tollerud}, {Ardelean}, {Babej}, {Bach}, {Bachetti}, {Bakanov}, {Bamford}, {Barentsen}, {Barmby}, {Baumbach}, {Berry}, {Biscani}, {Boquien}, {Bostroem}, {Bouma}, {Brammer}, {Bray}, {Breytenbach}, {Buddelmeijer}, {Burke}, {Calderone}, {Cano Rodr{\'\i}guez}, {Cara}, {Cardoso}, {Cheedella}, {Copin}, {Corrales}, {Crichton}, {D'Avella}, {Deil}, {Depagne}, {Dietrich}, {Donath}, {Droettboom}, {Earl}, {Erben}, {Fabbro}, {Ferreira}, {Finethy}, {Fox}, {Garrison}, {Gibbons}, {Goldstein}, {Gommers}, {Greco}, {Greenfield}, {Groener}, {Grollier}, {Hagen}, {Hirst}, {Homeier}, {Horton}, {Hosseinzadeh}, {Hu}, {Hunkeler}, {Ivezi{\'c}}, {Jain}, {Jenness}, {Kanarek}, {Kendrew}, {Kern}, {Kerzendorf}, {Khvalko}, {King}, {Kirkby}, {Kulkarni},
  {Kumar}, {Lee}, {Lenz}, {Littlefair}, {Ma}, {Macleod}, {Mastropietro}, {McCully}, {Montagnac}, {Morris}, {Mueller}, {Mumford}, {Muna}, {Murphy}, {Nelson}, {Nguyen}, {Ninan}, {N{\"o}the}, {Ogaz}, {Oh}, {Parejko}, {Parley}, {Pascual}, {Patil}, {Patil}, {Plunkett}, {Prochaska}, {Rastogi}, {Reddy Janga}, {Sabater}, {Sakurikar}, {Seifert}, {Sherbert}, {Sherwood-Taylor}, {Shih}, {Sick}, {Silbiger}, {Singanamalla}, {Singer}, {Sladen}, {Sooley}, {Sornarajah}, {Streicher}, {Teuben}, {Thomas}, {Tremblay}, {Turner}, {Terr{\'o}n}, {van Kerkwijk}, {de la Vega}, {Watkins}, {Weaver}, {Whitmore}, {Woillez}, {Zabalza}, \& {Astropy Contributors}}]{2018AJ....156..123A}
{Astropy Collaboration}, {Price-Whelan}, A.~M., {Sip{\H{o}}cz}, B.~M., {et~al.} 2018, \aj, 156, 123, \dodoi{10.3847/1538-3881/aabc4f}

\bibitem[{{Astropy Collaboration} {et~al.}(2022){Astropy Collaboration}, {Price-Whelan}, {Lim}, {Earl}, {Starkman}, {Bradley}, {Shupe}, {Patil}, {Corrales}, {Brasseur}, {N{\"o}the}, {Donath}, {Tollerud}, {Morris}, {Ginsburg}, {Vaher}, {Weaver}, {Tocknell}, {Jamieson}, {van Kerkwijk}, {Robitaille}, {Merry}, {Bachetti}, {G{\"u}nther}, {Aldcroft}, {Alvarado-Montes}, {Archibald}, {B{\'o}di}, {Bapat}, {Barentsen}, {Baz{\'a}n}, {Biswas}, {Boquien}, {Burke}, {Cara}, {Cara}, {Conroy}, {Conseil}, {Craig}, {Cross}, {Cruz}, {D'Eugenio}, {Dencheva}, {Devillepoix}, {Dietrich}, {Eigenbrot}, {Erben}, {Ferreira}, {Foreman-Mackey}, {Fox}, {Freij}, {Garg}, {Geda}, {Glattly}, {Gondhalekar}, {Gordon}, {Grant}, {Greenfield}, {Groener}, {Guest}, {Gurovich}, {Handberg}, {Hart}, {Hatfield-Dodds}, {Homeier}, {Hosseinzadeh}, {Jenness}, {Jones}, {Joseph}, {Kalmbach}, {Karamehmetoglu}, {Ka{\l}uszy{\'n}ski}, {Kelley}, {Kern}, {Kerzendorf}, {Koch}, {Kulumani}, {Lee}, {Ly}, {Ma}, {MacBride}, {Maljaars}, {Muna}, {Murphy}, {Norman},
  {O'Steen}, {Oman}, {Pacifici}, {Pascual}, {Pascual-Granado}, {Patil}, {Perren}, {Pickering}, {Rastogi}, {Roulston}, {Ryan}, {Rykoff}, {Sabater}, {Sakurikar}, {Salgado}, {Sanghi}, {Saunders}, {Savchenko}, {Schwardt}, {Seifert-Eckert}, {Shih}, {Jain}, {Shukla}, {Sick}, {Simpson}, {Singanamalla}, {Singer}, {Singhal}, {Sinha}, {Sip{\H{o}}cz}, {Spitler}, {Stansby}, {Streicher}, {{\v{S}}umak}, {Swinbank}, {Taranu}, {Tewary}, {Tremblay}, {de Val-Borro}, {Van Kooten}, {Vasovi{\'c}}, {Verma}, {de Miranda Cardoso}, {Williams}, {Wilson}, {Winkel}, {Wood-Vasey}, {Xue}, {Yoachim}, {Zhang}, {Zonca}, \& {Astropy Project Contributors}}]{2022ApJ...935..167A}
{Astropy Collaboration}, {Price-Whelan}, A.~M., {Lim}, P.~L., {et~al.} 2022, \apj, 935, 167, \dodoi{10.3847/1538-4357/ac7c74}

\bibitem[{{Bergvall} {et~al.}(2013){Bergvall}, {Leitet}, {Zackrisson}, \& {Marquart}}]{Bergvall2013}
{Bergvall}, N., {Leitet}, E., {Zackrisson}, E., \& {Marquart}, T. 2013, \aap, 554, A38, \dodoi{10.1051/0004-6361/201118433}

\bibitem[{{Bertin} \& {Arnouts}(1996)}]{1996A&AS..117..393B}
{Bertin}, E., \& {Arnouts}, S. 1996, \aaps, 117, 393, \dodoi{10.1051/aas:1996164}

\bibitem[{{Bridge} {et~al.}(2010){Bridge}, {Teplitz}, {Siana}, {Scarlata}, {Conselice}, {Ferguson}, {Brown}, {Salvato}, {Rudie}, {de Mello}, {Colbert}, {Gardner}, {Giavalisco}, \& {Armus}}]{Bridge2010}
{Bridge}, C.~R., {Teplitz}, H.~I., {Siana}, B., {et~al.} 2010, \apj, 720, 465, \dodoi{10.1088/0004-637X/720/1/465}

\bibitem[{{Conselice}(2003)}]{Conselice2003}
{Conselice}, C.~J. 2003, \apjs, 147, 1, \dodoi{10.1086/375001}

\bibitem[{{Dayal} {et~al.}(2024){Dayal}, {Volonteri}, {Greene}, {Kokorev}, {Goulding}, {Williams}, {Furtak}, {Zitrin}, {Atek}, {Chemerynska}, {Feldmann}, {Glazebrook}, {Labbe}, {Nanayakkara}, {Oesch}, \& {Weaver}}]{Dayal2024}
{Dayal}, P., {Volonteri}, M., {Greene}, J.~E., {et~al.} 2024, arXiv e-prints, arXiv:2401.11242, \dodoi{10.48550/arXiv.2401.11242}

\bibitem[{{de Barros} {et~al.}(2016){de Barros}, {Vanzella}, {Amor{\'\i}n}, {Castellano}, {Siana}, {Grazian}, {Suh}, {Balestra}, {Vignali}, {Verhamme}, {Zamorani}, {Mignoli}, {Hasinger}, {Comastri}, {Pentericci}, {P{\'e}rez-Montero}, {Fontana}, {Giavalisco}, \& {Gilli}}]{Debarros2016}
{de Barros}, S., {Vanzella}, E., {Amor{\'\i}n}, R., {et~al.} 2016, \aap, 585, A51, \dodoi{10.1051/0004-6361/201527046}

\bibitem[{{Dhiwar} {et~al.}(2024){Dhiwar}, {Saha}, {Maulick}, {Smith}, {Mondal}, {Teplitz}, {Rafelski}, \& {Windhorst}}]{Dhiwar2024}
{Dhiwar}, S., {Saha}, K., {Maulick}, S., {et~al.} 2024, \apjl, 963, L23, \dodoi{10.3847/2041-8213/ad2344}

\bibitem[{{Eisenstein} {et~al.}(2023{\natexlab{a}}){Eisenstein}, {Johnson}, {Robertson}, {Tacchella}, {Hainline}, {Jakobsen}, {Maiolino}, {Bonaventura}, {Bunker}, {Cameron}, {Cargile}, {Curtis-Lake}, {Hausen}, {Pusk{\'a}s}, {Rieke}, {Sun}, {Willmer}, {Willott}, {Alberts}, {Arribas}, {Baker}, {Baum}, {Bhatawdekar}, {Carniani}, {Charlot}, {Chen}, {Chevallard}, {Curti}, {DeCoursey}, {D'Eugenio}, {de Graaff}, {Egami}, {Helton}, {Ji}, {Jones}, {Kumari}, {L{\"u}tzgendorf}, {Laseter}, {Looser}, {Lyu}, {Maseda}, {Nelson}, {Parlanti}, {Rauscher}, {Rawle}, {Rieke}, {Rix}, {Rujopakarn}, {Sandles}, {Saxena}, {Scholtz}, {Sharpe}, {Shivaei}, {Simmonds}, {Smit}, {Topping}, {{\"U}bler}, {Venturi}, {Williams}, {Witstok}, \& {Woodrum}}]{Eisenstein2023b}
{Eisenstein}, D.~J., {Johnson}, B.~D., {Robertson}, B., {et~al.} 2023{\natexlab{a}}, arXiv e-prints, arXiv:2310.12340, \dodoi{10.48550/arXiv.2310.12340}

\bibitem[{{Eisenstein} {et~al.}(2023{\natexlab{b}}){Eisenstein}, {Willott}, {Alberts}, {Arribas}, {Bonaventura}, {Bunker}, {Cameron}, {Carniani}, {Charlot}, {Curtis-Lake}, {D'Eugenio}, {Endsley}, {Ferruit}, {Giardino}, {Hainline}, {Hausen}, {Jakobsen}, {Johnson}, {Maiolino}, {Rieke}, {Rieke}, {Rix}, {Robertson}, {Stark}, {Tacchella}, {Williams}, {Willmer}, {Baker}, {Baum}, {Bhatawdekar}, {Boyett}, {Chen}, {Chevallard}, {Circosta}, {Curti}, {Danhaive}, {DeCoursey}, {de Graaff}, {Dressler}, {Egami}, {Helton}, {Hviding}, {Ji}, {Jones}, {Kumari}, {L{\"u}tzgendorf}, {Laseter}, {Looser}, {Lyu}, {Maseda}, {Nelson}, {Parlanti}, {Perna}, {Pusk{\'a}s}, {Rawle}, {Rodr{\'\i}guez Del Pino}, {Sandles}, {Saxena}, {Scholtz}, {Sharpe}, {Shivaei}, {Silcock}, {Simmonds}, {Skarbinski}, {Smit}, {Stone}, {Suess}, {Sun}, {Tang}, {Topping}, {{\"U}bler}, {Villanueva}, {Wallace}, {Whitler}, {Witstok}, \& {Woodrum}}]{Eisenstein2023}
{Eisenstein}, D.~J., {Willott}, C., {Alberts}, S., {et~al.} 2023{\natexlab{b}}, arXiv e-prints, arXiv:2306.02465, \dodoi{10.48550/arXiv.2306.02465}

\bibitem[{{Fan} {et~al.}(2006){Fan}, {Strauss}, {Becker}, {White}, {Gunn}, {Knapp}, {Richards}, {Schneider}, {Brinkmann}, \& {Fukugita}}]{Fan2006}
{Fan}, X., {Strauss}, M.~A., {Becker}, R.~H., {et~al.} 2006, \aj, 132, 117, \dodoi{10.1086/504836}

\bibitem[{{Finkelstein} \& {Bagley}(2022)}]{Finkelstein2022}
{Finkelstein}, S.~L., \& {Bagley}, M.~B. 2022, \apj, 938, 25, \dodoi{10.3847/1538-4357/ac89eb}

\bibitem[{{Fletcher} {et~al.}(2019){Fletcher}, {Tang}, {Robertson}, {Nakajima}, {Ellis}, {Stark}, \& {Inoue}}]{Fletcher2019}
{Fletcher}, T.~J., {Tang}, M., {Robertson}, B.~E., {et~al.} 2019, \apj, 878, 87, \dodoi{10.3847/1538-4357/ab2045}

\bibitem[{{Flury} {et~al.}(2022){Flury}, {Jaskot}, {Ferguson}, {Worseck}, {Makan}, {Chisholm}, {Saldana-Lopez}, {Schaerer}, {McCandliss}, {Wang}, {Ford}, {Heckman}, {Ji}, {Giavalisco}, {Amorin}, {Atek}, {Blaizot}, {Borthakur}, {Carr}, {Castellano}, {Cristiani}, {De Barros}, {Dickinson}, {Finkelstein}, {Fleming}, {Fontanot}, {Garel}, {Grazian}, {Hayes}, {Henry}, {Mauerhofer}, {Micheva}, {Oey}, {Ostlin}, {Papovich}, {Pentericci}, {Ravindranath}, {Rosdahl}, {Rutkowski}, {Santini}, {Scarlata}, {Teplitz}, {Thuan}, {Trebitsch}, {Vanzella}, {Verhamme}, \& {Xu}}]{Flury2022}
{Flury}, S.~R., {Jaskot}, A.~E., {Ferguson}, H.~C., {et~al.} 2022, \apjs, 260, 1, \dodoi{10.3847/1538-4365/ac5331}

\bibitem[{{Gupta} {et~al.}(2024){Gupta}, {Trott}, {Jaiswar}, {Ryan-Weber}, {Bunker}, {Acharyya}, {Cameron}, {Forrest}, {Kacprzak}, {Nanayakkara}, {Tran}, \& {Chokshi}}]{Gupta2024}
{Gupta}, A., {Trott}, C.~M., {Jaiswar}, R., {et~al.} 2024, arXiv e-prints, arXiv:2403.13285, \dodoi{10.48550/arXiv.2403.13285}

\bibitem[{Harris {et~al.}(2020)Harris, Millman, van~der Walt, Gommers, Virtanen, Cournapeau, Wieser, Taylor, Berg, Smith, Kern, Picus, Hoyer, van Kerkwijk, Brett, Haldane, del R{\'{i}}o, Wiebe, Peterson, G{\'{e}}rard-Marchant, Sheppard, Reddy, Weckesser, Abbasi, Gohlke, \& Oliphant}]{harris2020array}
Harris, C.~R., Millman, K.~J., van~der Walt, S.~J., {et~al.} 2020, Nature, 585, 357, \dodoi{10.1038/s41586-020-2649-2}

\bibitem[{Hunter(2007)}]{Hunter:2007}
Hunter, J.~D. 2007, Computing in Science \& Engineering, 9, 90, \dodoi{10.1109/MCSE.2007.55}

\bibitem[{{Illingworth} {et~al.}(2016){Illingworth}, {Magee}, {Bouwens}, {Oesch}, {Labbe}, {van Dokkum}, {Whitaker}, {Holden}, {Franx}, \& {Gonzalez}}]{Illingworth2016}
{Illingworth}, G., {Magee}, D., {Bouwens}, R., {et~al.} 2016, arXiv e-prints, arXiv:1606.00841, \dodoi{10.48550/arXiv.1606.00841}

\bibitem[{{Inoue} {et~al.}(2014){Inoue}, {Shimizu}, {Iwata}, \& {Tanaka}}]{Inoue2014}
{Inoue}, A.~K., {Shimizu}, I., {Iwata}, I., \& {Tanaka}, M. 2014, \mnras, 442, 1805, \dodoi{10.1093/mnras/stu936}

\bibitem[{{Izotov} {et~al.}(2016{\natexlab{a}}){Izotov}, {Orlitov{\'a}}, {Schaerer}, {Thuan}, {Verhamme}, {Guseva}, \& {Worseck}}]{Izotov2016a}
{Izotov}, Y.~I., {Orlitov{\'a}}, I., {Schaerer}, D., {et~al.} 2016{\natexlab{a}}, \nat, 529, 178, \dodoi{10.1038/nature16456}

\bibitem[{{Izotov} {et~al.}(2016{\natexlab{b}}){Izotov}, {Schaerer}, {Thuan}, {Worseck}, {Guseva}, {Orlitov{\'a}}, \& {Verhamme}}]{Izotov2016b}
{Izotov}, Y.~I., {Schaerer}, D., {Thuan}, T.~X., {et~al.} 2016{\natexlab{b}}, \mnras, 461, 3683, \dodoi{10.1093/mnras/stw1205}

\bibitem[{{Izotov} {et~al.}(2018{\natexlab{a}}){Izotov}, {Schaerer}, {Worseck}, {Guseva}, {Thuan}, {Verhamme}, {Orlitov{\'a}}, \& {Fricke}}]{Izotov2018a}
{Izotov}, Y.~I., {Schaerer}, D., {Worseck}, G., {et~al.} 2018{\natexlab{a}}, \mnras, 474, 4514, \dodoi{10.1093/mnras/stx3115}

\bibitem[{{Izotov} {et~al.}(2021){Izotov}, {Worseck}, {Schaerer}, {Guseva}, {Chisholm}, {Thuan}, {Fricke}, \& {Verhamme}}]{Izotov2021}
{Izotov}, Y.~I., {Worseck}, G., {Schaerer}, D., {et~al.} 2021, \mnras, 503, 1734, \dodoi{10.1093/mnras/stab612}

\bibitem[{{Izotov} {et~al.}(2018{\natexlab{b}}){Izotov}, {Worseck}, {Schaerer}, {Guseva}, {Thuan}, {Fricke}, \& {Orlitov{\'a}}}]{Izotov2018b}
---. 2018{\natexlab{b}}, \mnras, 478, 4851, \dodoi{10.1093/mnras/sty1378}

\bibitem[{{Jaskot} \& {Oey}(2013)}]{Jaskot2013}
{Jaskot}, A.~E., \& {Oey}, M.~S. 2013, \apj, 766, 91, \dodoi{10.1088/0004-637X/766/2/91}

\bibitem[{{Ji} {et~al.}(2020){Ji}, {Giavalisco}, {Vanzella}, {Siana}, {Pentericci}, {Jaskot}, {Liu}, {Nonino}, {Ferguson}, {Castellano}, {Mannucci}, {Schaerer}, {Fynbo}, {Papovich}, {Carnall}, {Amorin}, {Simons}, {Hathi}, {Cullen}, \& {McLeod}}]{Ji2020}
{Ji}, Z., {Giavalisco}, M., {Vanzella}, E., {et~al.} 2020, \apj, 888, 109, \dodoi{10.3847/1538-4357/ab5fdc}

\bibitem[{{Jiang} {et~al.}(2013){Jiang}, {Egami}, {Fan}, {Windhorst}, {Cohen}, {Dav{\'e}}, {Finlator}, {Kashikawa}, {Mechtley}, {Ouchi}, \& {Shimasaku}}]{Jiang2013}
{Jiang}, L., {Egami}, E., {Fan}, X., {et~al.} 2013, \apj, 773, 153, \dodoi{10.1088/0004-637X/773/2/153}

\bibitem[{{Jiang} {et~al.}(2022){Jiang}, {Ning}, {Fan}, {Ho}, {Luo}, {Wang}, {Wu}, {Wu}, {Yang}, \& {Zheng}}]{Jiang2022}
{Jiang}, L., {Ning}, Y., {Fan}, X., {et~al.} 2022, Nature Astronomy, 6, 850, \dodoi{10.1038/s41550-022-01708-w}

\bibitem[{{Kartaltepe} {et~al.}(2010){Kartaltepe}, {Sanders}, {Le Floc'h}, {Frayer}, {Aussel}, {Arnouts}, {Ilbert}, {Salvato}, {Scoville}, {Surace}, {Yan}, {Capak}, {Caputi}, {Carollo}, {Cassata}, {Civano}, {Hasinger}, {Koekemoer}, {Le F{\`e}vre}, {Lilly}, {Liu}, {McCracken}, {Schinnerer}, {Smol{\v{c}}i{\'c}}, {Taniguchi}, {Thompson}, {Trump}, {Baldassare}, \& {Fiorenza}}]{Kartaltepe2010}
{Kartaltepe}, J.~S., {Sanders}, D.~B., {Le Floc'h}, E., {et~al.} 2010, \apj, 721, 98, \dodoi{10.1088/0004-637X/721/1/98}

\bibitem[{{Kawamata} {et~al.}(2018){Kawamata}, {Ishigaki}, {Shimasaku}, {Oguri}, {Ouchi}, \& {Tanigawa}}]{Kawamata2018}
{Kawamata}, R., {Ishigaki}, M., {Shimasaku}, K., {et~al.} 2018, \apj, 855, 4, \dodoi{10.3847/1538-4357/aaa6cf}

\bibitem[{{Kerutt} {et~al.}(2024){Kerutt}, {Oesch}, {Wisotzki}, {Verhamme}, {Atek}, {Herenz}, {Illingworth}, {Kusakabe}, {Matthee}, {Mauerhofer}, {Montes}, {Naidu}, {Nelson}, {Reddy}, {Schaye}, {Simmonds}, {Urrutia}, \& {Vitte}}]{Kerutt2024}
{Kerutt}, J., {Oesch}, P.~A., {Wisotzki}, L., {et~al.} 2024, \aap, 684, A42, \dodoi{10.1051/0004-6361/202346656}

\bibitem[{{Le Reste} {et~al.}(2023){Le Reste}, {Cannon}, {Hayes}, {Inoue}, {Kepley}, {Melinder}, {Menacho}, {Adamo}, {Bik}, {Ejdetj{\"a}rn}, {J{\'o}zsa}, {{\"O}stlin}, \& {Taft}}]{LeReste2023}
{Le Reste}, A., {Cannon}, J.~M., {Hayes}, M.~J., {et~al.} 2023, \mnras, \dodoi{10.1093/mnras/stad3910}

\bibitem[{{Leclercq} {et~al.}(2024){Leclercq}, {Chisholm}, {King}, {Zeimann}, {Jaskot}, {Henry}, {Hayes}, {Flury}, {Izotov}, {Prochaska}, {Verhamme}, {Amor{\'\i}n}, {Atek}, {Bait}, {Blaizot}, {Carr}, {Ji}, {Le Reste}, {Ferguson}, {Gazagnes}, {Heckman}, {Komarova}, {Marques-Chaves}, {{\"O}stlin}, {Saldana-Lopez}, {Scarlata}, {Schaerer}, {Thuan}, {Trebitsch}, {Worseck}, {Wang}, \& {Xu}}]{Leclercq2024}
{Leclercq}, F., {Chisholm}, J., {King}, W., {et~al.} 2024, arXiv e-prints, arXiv:2401.14981, \dodoi{10.48550/arXiv.2401.14981}

\bibitem[{{Lotz} {et~al.}(2008){Lotz}, {Jonsson}, {Cox}, \& {Primack}}]{Lotz2008b}
{Lotz}, J.~M., {Jonsson}, P., {Cox}, T.~J., \& {Primack}, J.~R. 2008, \mnras, 391, 1137, \dodoi{10.1111/j.1365-2966.2008.14004.x}

\bibitem[{{Lotz} {et~al.}(2004){Lotz}, {Primack}, \& {Madau}}]{Lotz2004}
{Lotz}, J.~M., {Primack}, J., \& {Madau}, P. 2004, \aj, 128, 163, \dodoi{10.1086/421849}

\bibitem[{{Madau} \& {Haardt}(2015)}]{Madau2015}
{Madau}, P., \& {Haardt}, F. 2015, \apjl, 813, L8, \dodoi{10.1088/2041-8205/813/1/L8}

\bibitem[{{Malhotra} {et~al.}(2012){Malhotra}, {Rhoads}, {Finkelstein}, {Hathi}, {Nilsson}, {McLinden}, \& {Pirzkal}}]{Malhotra2012}
{Malhotra}, S., {Rhoads}, J.~E., {Finkelstein}, S.~L., {et~al.} 2012, \apjl, 750, L36, \dodoi{10.1088/2041-8205/750/2/L36}

\bibitem[{{Marques-Chaves} {et~al.}(2021){Marques-Chaves}, {Schaerer}, {{\'A}lvarez-M{\'a}rquez}, {Colina}, {Dessauges-Zavadsky}, {P{\'e}rez-Fournon}, {Saldana-Lopez}, \& {Verhamme}}]{Marques-Chaves2021}
{Marques-Chaves}, R., {Schaerer}, D., {{\'A}lvarez-M{\'a}rquez}, J., {et~al.} 2021, \mnras, 507, 524, \dodoi{10.1093/mnras/stab2187}

\bibitem[{{Marques-Chaves} {et~al.}(2022){Marques-Chaves}, {Schaerer}, {{\'A}lvarez-M{\'a}rquez}, {Verhamme}, {Ceverino}, {Chisholm}, {Colina}, {Dessauges-Zavadsky}, {P{\'e}rez-Fournon}, {Saldana-Lopez}, {Upadhyaya}, \& {Vanzella}}]{Marques-Chaves2022}
---. 2022, \mnras, 517, 2972, \dodoi{10.1093/mnras/stac2893}

\bibitem[{{Marques-Chaves} {et~al.}(2024){Marques-Chaves}, {Schaerer}, {Vanzella}, {Verhamme}, {Dessauges-Zavadsky}, {Chisholm}, {Leclercq}, {Upadhyaya}, {Alvarez-Marquez}, {Colina}, {Garel}, \& {Messa}}]{Marques-Chaves2024}
{Marques-Chaves}, R., {Schaerer}, D., {Vanzella}, E., {et~al.} 2024, arXiv e-prints, arXiv:2407.18804, \dodoi{10.48550/arXiv.2407.18804}

\bibitem[{{Matsuoka} {et~al.}(2023){Matsuoka}, {Onoue}, {Iwasawa}, {Strauss}, {Kashikawa}, {Izumi}, {Nagao}, {Imanishi}, {Akiyama}, {Silverman}, {Asami}, {Bosch}, {Furusawa}, {Goto}, {Gunn}, {Harikane}, {Ikeda}, {Inayoshi}, {Ishimoto}, {Kawaguchi}, {Kikuta}, {Kohno}, {Komiyama}, {Lee}, {Lupton}, {Minezaki}, {Miyazaki}, {Murayama}, {Nishizawa}, {Oguri}, {Ono}, {Oogi}, {Ouchi}, {Price}, {Sameshima}, {Sugiyama}, {Tait}, {Takada}, {Takahashi}, {Takata}, {Tanaka}, {Toba}, {Wang}, \& {Yamashita}}]{Matsuoka2023}
{Matsuoka}, Y., {Onoue}, M., {Iwasawa}, K., {et~al.} 2023, \apjl, 949, L42, \dodoi{10.3847/2041-8213/acd69f}

\bibitem[{{Matthee} {et~al.}(2024){Matthee}, {Naidu}, {Brammer}, {Chisholm}, {Eilers}, {Goulding}, {Greene}, {Kashino}, {Labbe}, {Lilly}, {Mackenzie}, {Oesch}, {Weibel}, {Wuyts}, {Xiao}, {Bordoloi}, {Bouwens}, {van Dokkum}, {Illingworth}, {Kramarenko}, {Maseda}, {Mason}, {Meyer}, {Nelson}, {Reddy}, {Shivaei}, {Simcoe}, \& {Yue}}]{Matthee2024}
{Matthee}, J., {Naidu}, R.~P., {Brammer}, G., {et~al.} 2024, \apj, 963, 129, \dodoi{10.3847/1538-4357/ad2345}

\bibitem[{{Maulick} {et~al.}(2024{\natexlab{a}}){Maulick}, {Saha}, {Kataria}, \& {Herenz}}]{Maulick2024a}
{Maulick}, S., {Saha}, K., {Kataria}, M., \& {Herenz}, E.~C. 2024{\natexlab{a}}, \apj, 972, 138, \dodoi{10.3847/1538-4357/ad6155}

\bibitem[{{Maulick} {et~al.}(2024{\natexlab{b}}){Maulick}, {Saha}, \& {Rutkowski}}]{Maulick2024b}
{Maulick}, S., {Saha}, K., \& {Rutkowski}, M.~J. 2024{\natexlab{b}}, arXiv e-prints, arXiv:2410.09515, \dodoi{10.48550/arXiv.2410.09515}

\bibitem[{Momcheva(2017)}]{https://doi.org/10.17909/t9jw9z}
Momcheva, I. 2017, 3D-HST,  STScI/MAST, \dodoi{10.17909/T9JW9Z}

\bibitem[{{Morishita} {et~al.}(2024){Morishita}, {Stiavelli}, {Chary}, {Trenti}, {Bergamini}, {Chiaberge}, {Leethochawalit}, {Roberts-Borsani}, {Shen}, \& {Treu}}]{Morishita2024}
{Morishita}, T., {Stiavelli}, M., {Chary}, R.-R., {et~al.} 2024, \apj, 963, 9, \dodoi{10.3847/1538-4357/ad1404}

\bibitem[{{Naidu} {et~al.}(2020){Naidu}, {Tacchella}, {Mason}, {Bose}, {Oesch}, \& {Conroy}}]{Naidu2020}
{Naidu}, R.~P., {Tacchella}, S., {Mason}, C.~A., {et~al.} 2020, \apj, 892, 109, \dodoi{10.3847/1538-4357/ab7cc9}

\bibitem[{{Nakajima} \& {Ouchi}(2014)}]{Nakajima2014}
{Nakajima}, K., \& {Ouchi}, M. 2014, \mnras, 442, 900, \dodoi{10.1093/mnras/stu902}

\bibitem[{{Oesch} {et~al.}(2018){Oesch}, {Montes}, {Reddy}, {Bouwens}, {Illingworth}, {Magee}, {Atek}, {Carollo}, {Cibinel}, {Franx}, {Holden}, {Labb{\'e}}, {Nelson}, {Steidel}, {van Dokkum}, {Morselli}, {Naidu}, \& {Wilkins}}]{Oesch2018}
{Oesch}, P.~A., {Montes}, M., {Reddy}, N., {et~al.} 2018, \apjs, 237, 12, \dodoi{10.3847/1538-4365/aacb30}

\bibitem[{{Oesch} {et~al.}(2023){Oesch}, {Brammer}, {Naidu}, {Bouwens}, {Chisholm}, {Illingworth}, {Matthee}, {Nelson}, {Qin}, {Reddy}, {Shapley}, {Shivaei}, {van Dokkum}, {Weibel}, {Whitaker}, {Wuyts}, {Covelo-Paz}, {Endsley}, {Fudamoto}, {Giovinazzo}, {Herard-Demanche}, {Kerutt}, {Kramarenko}, {Labbe}, {Leonova}, {Lin}, {Magee}, {Marchesini}, {Maseda}, {Mason}, {Matharu}, {Meyer}, {Neufeld}, {Prieto Lyon}, {Schaerer}, {Sharma}, {Shuntov}, {Smit}, {Stefanon}, {Wyithe}, \& {Xiao}}]{Oesch2023}
{Oesch}, P.~A., {Brammer}, G., {Naidu}, R.~P., {et~al.} 2023, \mnras, 525, 2864, \dodoi{10.1093/mnras/stad2411}

\bibitem[{{Peng} {et~al.}(2010){Peng}, {Ho}, {Impey}, \& {Rix}}]{Peng2010}
{Peng}, C.~Y., {Ho}, L.~C., {Impey}, C.~D., \& {Rix}, H.-W. 2010, \aj, 139, 2097, \dodoi{10.1088/0004-6256/139/6/2097}

\bibitem[{{Planck Collaboration} {et~al.}(2020){Planck Collaboration}, {Aghanim}, {Akrami}, {Arroja}, {Ashdown}, {Aumont}, {Baccigalupi}, {Ballardini}, {Banday}, {Barreiro}, {Bartolo}, {Basak}, {Battye}, {Benabed}, {Bernard}, {Bersanelli}, {Bielewicz}, {Bock}, {Bond}, {Borrill}, {Bouchet}, {Boulanger}, {Bucher}, {Burigana}, {Butler}, {Calabrese}, {Cardoso}, {Carron}, {Casaponsa}, {Challinor}, {Chiang}, {Colombo}, {Combet}, {Contreras}, {Crill}, {Cuttaia}, {de Bernardis}, {de Zotti}, {Delabrouille}, {Delouis}, {D{\'e}sert}, {Di Valentino}, {Dickinson}, {Diego}, {Donzelli}, {Dor{\'e}}, {Douspis}, {Ducout}, {Dupac}, {Efstathiou}, {Elsner}, {En{\ss}lin}, {Eriksen}, {Falgarone}, {Fantaye}, {Fergusson}, {Fernandez-Cobos}, {Finelli}, {Forastieri}, {Frailis}, {Franceschi}, {Frolov}, {Galeotta}, {Galli}, {Ganga}, {G{\'e}nova-Santos}, {Gerbino}, {Ghosh}, {Gonz{\'a}lez-Nuevo}, {G{\'o}rski}, {Gratton}, {Gruppuso}, {Gudmundsson}, {Hamann}, {Handley}, {Hansen}, {Helou}, {Herranz}, {Hildebrandt}, {Hivon}, {Huang}, {Jaffe},
  {Jones}, {Karakci}, {Keih{\"a}nen}, {Keskitalo}, {Kiiveri}, {Kim}, {Kisner}, {Knox}, {Krachmalnicoff}, {Kunz}, {Kurki-Suonio}, {Lagache}, {Lamarre}, {Langer}, {Lasenby}, {Lattanzi}, {Lawrence}, {Le Jeune}, {Leahy}, {Lesgourgues}, {Levrier}, {Lewis}, {Liguori}, {Lilje}, {Lilley}, {Lindholm}, {L{\'o}pez-Caniego}, {Lubin}, {Ma}, {Mac{\'\i}as-P{\'e}rez}, {Maggio}, {Maino}, {Mandolesi}, {Mangilli}, {Marcos-Caballero}, {Maris}, {Martin}, {Martinelli}, {Mart{\'\i}nez-Gonz{\'a}lez}, {Matarrese}, {Mauri}, {McEwen}, {Meerburg}, {Meinhold}, {Melchiorri}, {Mennella}, {Migliaccio}, {Millea}, {Mitra}, {Miville-Desch{\^e}nes}, {Molinari}, {Moneti}, {Montier}, {Morgante}, {Moss}, {Mottet}, {M{\"u}nchmeyer}, {Natoli}, {N{\o}rgaard-Nielsen}, {Oxborrow}, {Pagano}, {Paoletti}, {Partridge}, {Patanchon}, {Pearson}, {Peel}, {Peiris}, {Perrotta}, {Pettorino}, {Piacentini}, {Polastri}, {Polenta}, {Puget}, {Rachen}, {Reinecke}, {Remazeilles}, {Renault}, {Renzi}, {Rocha}, {Rosset}, {Roudier}, {Rubi{\~n}o-Mart{\'\i}n},
  {Ruiz-Granados}, {Salvati}, {Sandri}, {Savelainen}, {Scott}, {Shellard}, {Shiraishi}, {Sirignano}, {Sirri}, {Spencer}, {Sunyaev}, {Suur-Uski}, {Tauber}, {Tavagnacco}, {Tenti}, {Terenzi}, {Toffolatti}, {Tomasi}, {Trombetti}, {Valiviita}, {Van Tent}, {Vibert}, {Vielva}, {Villa}, {Vittorio}, {Wandelt}, {Wehus}, {White}, {White}, {Zacchei}, \& {Zonca}}]{Planck2020}
{Planck Collaboration}, {Aghanim}, N., {Akrami}, Y., {et~al.} 2020, \aap, 641, A1, \dodoi{10.1051/0004-6361/201833880}

\bibitem[{{Rauch} {et~al.}(2011){Rauch}, {Becker}, {Haehnelt}, {Gauthier}, {Ravindranath}, \& {Sargent}}]{Rauch2011}
{Rauch}, M., {Becker}, G.~D., {Haehnelt}, M.~G., {et~al.} 2011, \mnras, 418, 1115, \dodoi{10.1111/j.1365-2966.2011.19556.x}

\bibitem[{{Rieke, Marcia} {et~al.}(2023){Rieke, Marcia}, {Robertson, Brant}, {Tacchella, Sandro}, {Willmer, Christopher}, {Johnson, Ben}, {Carniani, Stefano}, {Bunker, Andy}, \& {Willott, Chris}}]{https://doi.org/10.17909/8tdj-8n28}
{Rieke, Marcia}, {Robertson, Brant}, {Tacchella, Sandro}, {et~al.} 2023, Data from the JWST Advanced Deep Extragalactic Survey (JADES),  STScI/MAST, \dodoi{10.17909/8TDJ-8N28}

\bibitem[{{Rivera-Thorsen} {et~al.}(2022){Rivera-Thorsen}, {Hayes}, \& {Melinder}}]{Rivera-thorsen2022}
{Rivera-Thorsen}, T.~E., {Hayes}, M., \& {Melinder}, J. 2022, arXiv e-prints, arXiv:2206.10799.
\newblock \doarXiv{2206.10799}

\bibitem[{{Robertson}(2022)}]{Robertson2022}
{Robertson}, B.~E. 2022, \araa, 60, 121, \dodoi{10.1146/annurev-astro-120221-044656}

\bibitem[{{Robertson} {et~al.}(2015){Robertson}, {Ellis}, {Furlanetto}, \& {Dunlop}}]{Robertson2015}
{Robertson}, B.~E., {Ellis}, R.~S., {Furlanetto}, S.~R., \& {Dunlop}, J.~S. 2015, \apjl, 802, L19, \dodoi{10.1088/2041-8205/802/2/L19}

\bibitem[{{Rodriguez-Gomez} {et~al.}(2019){Rodriguez-Gomez}, {Snyder}, {Lotz}, {Nelson}, {Pillepich}, {Springel}, {Genel}, {Weinberger}, {Tacchella}, {Pakmor}, {Torrey}, {Marinacci}, {Vogelsberger}, {Hernquist}, \& {Thilker}}]{2019MNRAS.483.4140R}
{Rodriguez-Gomez}, V., {Snyder}, G.~F., {Lotz}, J.~M., {et~al.} 2019, \mnras, 483, 4140, \dodoi{10.1093/mnras/sty3345}

\bibitem[{{Rose} {et~al.}(2023){Rose}, {Kartaltepe}, {Snyder}, {Rodriguez-Gomez}, {Yung}, {Arrabal Haro}, {Bagley}, {Calabr{\'o}}, {Cleri}, {Cooper}, {Costantin}, {Croton}, {Dickinson}, {Finkelstein}, {H{\"a}u{\ss}ler}, {Holwerda}, {Koekemoer}, {Kurczynski}, {Lucas}, {Mantha}, {Papovich}, {P{\'e}rez-Gonz{\'a}lez}, {Pirzkal}, {Somerville}, {Straughn}, \& {Tacchella}}]{Rose2023}
{Rose}, C., {Kartaltepe}, J.~S., {Snyder}, G.~F., {et~al.} 2023, \apj, 942, 54, \dodoi{10.3847/1538-4357/ac9f10}

\bibitem[{{Roy} {et~al.}(2024){Roy}, {Heckman}, {Henry}, {Chisholm}, {Flury}, {Leitherer}, {Hayes}, {Jaskot}, {Ji}, {Schaerer}, {Wang}, {Borthakur}, {Xu}, \& {{\"O}stlin}}]{Roy2024}
{Roy}, N., {Heckman}, T., {Henry}, A., {et~al.} 2024, arXiv e-prints, arXiv:2410.13254, \dodoi{10.48550/arXiv.2410.13254}

\bibitem[{{Saha} {et~al.}(2020){Saha}, {Tandon}, {Simmonds}, {Verhamme}, {Paswan}, {Schaerer}, {Rutkowski}, {Borgohain}, {Elmegreen}, {Inoue}, {Combes}, {Elmegreen}, \& {Paalvast}}]{Saha2020}
{Saha}, K., {Tandon}, S.~N., {Simmonds}, C., {et~al.} 2020, Nature Astronomy, 4, 1185, \dodoi{10.1038/s41550-020-1173-5}

\bibitem[{{Saxena} {et~al.}(2022){Saxena}, {Pentericci}, {Ellis}, {Guaita}, {Calabr{\`o}}, {Schaerer}, {Vanzella}, {Amor{\'\i}n}, {Bolzonella}, {Castellano}, {Fontanot}, {Hathi}, {Hibon}, {Llerena}, {Mannucci}, {Saldana-Lopez}, {Talia}, \& {Zamorani}}]{Saxena2022}
{Saxena}, A., {Pentericci}, L., {Ellis}, R.~S., {et~al.} 2022, \mnras, 511, 120, \dodoi{10.1093/mnras/stab3728}

\bibitem[{{Shapley} {et~al.}(2016){Shapley}, {Steidel}, {Strom}, {Bogosavljevi{\'c}}, {Reddy}, {Siana}, {Mostardi}, \& {Rudie}}]{Shapley2016}
{Shapley}, A.~E., {Steidel}, C.~C., {Strom}, A.~L., {et~al.} 2016, \apjl, 826, L24, \dodoi{10.3847/2041-8205/826/2/L24}

\bibitem[{{Shibuya} {et~al.}(2015){Shibuya}, {Ouchi}, \& {Harikane}}]{Shibuya2015}
{Shibuya}, T., {Ouchi}, M., \& {Harikane}, Y. 2015, \apjs, 219, 15, \dodoi{10.1088/0067-0049/219/2/15}

\bibitem[{{Snyder} {et~al.}(2019){Snyder}, {Rodriguez-Gomez}, {Lotz}, {Torrey}, {Quirk}, {Hernquist}, {Vogelsberger}, \& {Freeman}}]{Snyder2019}
{Snyder}, G.~F., {Rodriguez-Gomez}, V., {Lotz}, J.~M., {et~al.} 2019, \mnras, 486, 3702, \dodoi{10.1093/mnras/stz1059}

\bibitem[{{Steidel} {et~al.}(2018){Steidel}, {Bogosavljevi{\'c}}, {Shapley}, {Reddy}, {Rudie}, {Pettini}, {Trainor}, \& {Strom}}]{Steidel2018}
{Steidel}, C.~C., {Bogosavljevi{\'c}}, M., {Shapley}, A.~E., {et~al.} 2018, \apj, 869, 123, \dodoi{10.3847/1538-4357/aaed28}

\bibitem[{{Vanzella} {et~al.}(2012){Vanzella}, {Guo}, {Giavalisco}, {Grazian}, {Castellano}, {Cristiani}, {Dickinson}, {Fontana}, {Nonino}, {Giallongo}, {Pentericci}, {Galametz}, {Faber}, {Ferguson}, {Grogin}, {Koekemoer}, {Newman}, \& {Siana}}]{Vanzella2012}
{Vanzella}, E., {Guo}, Y., {Giavalisco}, M., {et~al.} 2012, \apj, 751, 70, \dodoi{10.1088/0004-637X/751/1/70}

\bibitem[{{Vanzella} {et~al.}(2016){Vanzella}, {de Barros}, {Vasei}, {Alavi}, {Giavalisco}, {Siana}, {Grazian}, {Hasinger}, {Suh}, {Cappelluti}, {Vito}, {Amorin}, {Balestra}, {Brusa}, {Calura}, {Castellano}, {Comastri}, {Fontana}, {Gilli}, {Mignoli}, {Pentericci}, {Vignali}, \& {Zamorani}}]{Vanzella2016}
{Vanzella}, E., {de Barros}, S., {Vasei}, K., {et~al.} 2016, \apj, 825, 41, \dodoi{10.3847/0004-637X/825/1/41}

\bibitem[{{Vanzella} {et~al.}(2018){Vanzella}, {Nonino}, {Cupani}, {Castellano}, {Sani}, {Mignoli}, {Calura}, {Meneghetti}, {Gilli}, {Comastri}, {Mercurio}, {Caminha}, {Caputi}, {Rosati}, {Grillo}, {Cristiani}, {Balestra}, {Fontana}, \& {Giavalisco}}]{Vanzella2018}
{Vanzella}, E., {Nonino}, M., {Cupani}, G., {et~al.} 2018, \mnras, 476, L15, \dodoi{10.1093/mnrasl/sly023}

\bibitem[{{Verhamme} {et~al.}(2015){Verhamme}, {Orlitov{\'a}}, {Schaerer}, \& {Hayes}}]{Verhamme2015}
{Verhamme}, A., {Orlitov{\'a}}, I., {Schaerer}, D., \& {Hayes}, M. 2015, \aap, 578, A7, \dodoi{10.1051/0004-6361/201423978}

\bibitem[{{Verhamme} {et~al.}(2017){Verhamme}, {Orlitov{\'a}}, {Schaerer}, {Izotov}, {Worseck}, {Thuan}, \& {Guseva}}]{Verhamme2017}
{Verhamme}, A., {Orlitov{\'a}}, I., {Schaerer}, D., {et~al.} 2017, \aap, 597, A13, \dodoi{10.1051/0004-6361/201629264}

\bibitem[{{Wang} {et~al.}(2019){Wang}, {Heckman}, {Leitherer}, {Alexandroff}, {Borthakur}, \& {Overzier}}]{Wang2019}
{Wang}, B., {Heckman}, T.~M., {Leitherer}, C., {et~al.} 2019, \apj, 885, 57, \dodoi{10.3847/1538-4357/ab418f}

\bibitem[{{Whitaker} {et~al.}(2019){Whitaker}, {Ashas}, {Illingworth}, {Magee}, {Leja}, {Oesch}, {van Dokkum}, {Mowla}, {Bouwens}, {Franx}, {Holden}, {Labb{\'e}}, {Rafelski}, {Teplitz}, \& {Gonzalez}}]{Whitaker2019}
{Whitaker}, K.~E., {Ashas}, M., {Illingworth}, G., {et~al.} 2019, \apjs, 244, 16, \dodoi{10.3847/1538-4365/ab3853}

\bibitem[{{Williams} {et~al.}(2023){Williams}, {Tacchella}, {Maseda}, {Robertson}, {Johnson}, {Willott}, {Eisenstein}, {Willmer}, {Ji}, {Hainline}, {Helton}, {Alberts}, {Baum}, {Bhatawdekar}, {Boyett}, {Bunker}, {Carniani}, {Charlot}, {Chevallard}, {Curtis-Lake}, {de Graaff}, {Egami}, {Franx}, {Kumari}, {Maiolino}, {Nelson}, {Rieke}, {Sandles}, {Shivaei}, {Simmonds}, {Smit}, {Suess}, {Sun}, {{\"U}bler}, \& {Witstok}}]{Williams2023}
{Williams}, C.~C., {Tacchella}, S., {Maseda}, M.~V., {et~al.} 2023, \apjs, 268, 64, \dodoi{10.3847/1538-4365/acf130}

\bibitem[{{Williams, Christina} {et~al.}(2023){Williams, Christina}, {Tacchella, Sandro}, \& {Maseda, Michael}}]{https://doi.org/10.17909/fsc4-dt61}
{Williams, Christina}, {Tacchella, Sandro}, \& {Maseda, Michael}. 2023, Data from the JWST Extragalactic Medium-band Survey (JEMS),  STScI/MAST, \dodoi{10.17909/FSC4-DT61}

\bibitem[{{Witten} {et~al.}(2024){Witten}, {Laporte}, {Martin-Alvarez}, {Sijacki}, {Yuan}, {Haehnelt}, {Baker}, {Dunlop}, {Ellis}, {Grogin}, {Illingworth}, {Katz}, {Koekemoer}, {Magee}, {Maiolino}, {McClymont}, {P{\'e}rez-Gonz{\'a}lez}, {Pusk{\'a}s}, {Roberts-Borsani}, {Santini}, \& {Simmonds}}]{Witten2024}
{Witten}, C., {Laporte}, N., {Martin-Alvarez}, S., {et~al.} 2024, Nature Astronomy, 8, 384, \dodoi{10.1038/s41550-023-02179-3}

\bibitem[{{Yuan} {et~al.}(2024){Yuan}, {Zheng}, {Jiang}, {Zhu}, {Lin}, \& {Cheng}}]{Yuan2024}
{Yuan}, F.-T., {Zheng}, Z.-Y., {Jiang}, C., {et~al.} 2024, arXiv e-prints, arXiv:2409.20352, \dodoi{10.48550/arXiv.2409.20352}

\bibitem[{{Yuan} {et~al.}(2021){Yuan}, {Zheng}, {Lin}, {Zhu}, \& {Rahna}}]{Yuan2021}
{Yuan}, F.-T., {Zheng}, Z.-Y., {Lin}, R., {Zhu}, S., \& {Rahna}, P.~T. 2021, \apjl, 923, L28, \dodoi{10.3847/2041-8213/ac4170}

\bibitem[{{Zhan}(2018)}]{zhan2018}
{Zhan}, H. 2018, in 42nd COSPAR Scientific Assembly, Vol.~42, E1.16--4--18

\bibitem[{{Zhu} {et~al.}(2024){Zhu}, {Yuan}, {Jiang}, {Zheng}, \& {Lin}}]{Zhu2024}
{Zhu}, S., {Yuan}, F.-T., {Jiang}, C., {Zheng}, Z.-Y., \& {Lin}, R. 2024, arXiv e-prints, arXiv:2409.20349, \dodoi{10.48550/arXiv.2409.20349}

\end{thebibliography}
\bibliographystyle{aasjournal}
\end{CJK*}
\end{document}